\documentclass[preprintnumbers,10pt,nofootinbib]{revtex4}

\usepackage{amsmath,latexsym,amssymb,amsfonts}
\usepackage[dvips]{color,graphicx}
\begin{document}

\preprint{IPMU18-0102, YITP-18-60}

\title{\textbf{Hawking radiation as instantons}}
\author{
\textsc{Pisin Chen$^{a,b,c,d}$}\footnote{{\tt pisinchen{}@{}phys.ntu.edu.tw}},
\textsc{Misao Sasaki$^{a,e,f}$}\footnote{{\tt misao.sasaki{}@{}ipmu.jp}} 
and
\textsc{Dong-han Yeom$^{g,h,i,j}$}\footnote{{\tt innocent.yeom{}@{}gmail.com}}
}

\affiliation{
$^{a}$\small{Leung Center for Cosmology and Particle Astrophysics, National Taiwan University, Taipei 10617, Taiwan}\\
$^{b}$\small{Department of Physics, National Taiwan University, Taipei 10617, Taiwan}\\
$^{c}$\small{Graduate Institute of Astrophysics, National Taiwan University, Taipei 10617, Taiwan}\\
$^{d}$\small{Kavli Institute for Particle Astrophysics and Cosmology,
SLAC National Accelerator Laboratory, Stanford University, Stanford, California 94305, USA}\\
$^{e}$\small{Kavli Institute for the Physics and Mathematics of the Universe (WPI), 
	University of Tokyo, Chiba 277-8583, Japan}\\
$^{f}$\small{Yukawa Institute for Theoretical Physics, Kyoto University, Kyoto 606-8502, Japan}\\
$^{g}$\small{Asia Pacific Center for Theoretical Physics, Pohang 37673, Republic of Korea}\\
$^{h}$\small{Department of Physics, POSTECH, Pohang 37673, Republic of Korea}\\
$^{i}$\small{Department of Physics Education, Pusan National University, Busan 46241, Republic of Korea}\\
$^{j}$\small{Research Center for Dielectric and Advanced Matter Physics, Pusan National University, Busan 46241, Republic of Korea}
}

\begin{abstract}
There have been various interpretations of Hawking radiation proposed based on the perturbative approach, and all have confirmed Hawking's original finding. One major conceptual challenge of Hawking evaporation is the associated black hole information loss paradox, which remains unresolved. A key factor to the issue is the end-stage of  the black hole evaporation. Unfortunately by then the evaporation process becomes non-perturbative. Aspired to provide a tool for the eventual solution to this problem, here we introduce a new interpretation of Hawking radiation as the tunneling of instantons. 
We study instantons of a massless scalar field in Einstein gravity. 
We consider a complex-valued instanton that connects an initial pure black hole state to a black hole with
a scalar field that represents the Hawking radiation at future null infinity, where its action depends only on the areal entropy difference. 
By comparing it with several independent approaches to Hawking radiation in the perturbative limit, we conclude that 
Hawking radiation may indeed be described by a family of instantons. Since the instanton approach can describe non-perturbative processes, we hope that our new interpretation and holistic method may shed lights on the information loss problem.
\end{abstract}

\maketitle

\newpage

\tableofcontents

%\newpage

\section{Introduction}

One of the most challenging issues in black hole physics is the information loss problem. A black hole evaporates via
Hawking radiation~\cite{Hawking:1974sw}. The evaporation depends only on the property of the horizon and
hence after the evaporation, it seems that the original information cannot be recovered~\cite{Hawking:1976ra}. 
The problem has been addressed from various points of view, yet it remains unresolved.
One aspect of the problem that may be agreed by most experts is the preservation of unitarity, which is also 
supported by the notion of AdS/CFT correspondence \cite{Maldacena:1997re}. Many ideas have been proposed to explain the unitarity maybe preserved during the black hole evaporation (e.g., \cite{Susskind:1993if,Almheiri:2012rt}), but no consensus has been reached yet (e.g., \cite{Chen:2014jwq,Chen:2015gux}).
A conventional interpretation is the black hole complementarity conjecture \cite{Susskind:1993if}, according to which information is recovered by Hawking radiation through quantum entanglement \cite{Page:1993wv} 
 without inconsistency. However, some counterexamples and criticisms have recently been reported by 
 several authors \cite{Yeom:2008qw,Hwang:2012nn,Almheiri:2012rt}, but the AMPS firewall conjecture \cite{Almheiri:2012rt} is also challenged \cite{Chen:2015gux}.
 
One difficulty is that there exist only limited tools at hand to deal with the physics of black hole evaporation.
If one only relies on perturbative quantum field theory (e.g., \cite{Callan:1992rs}), then the information loss is apparent. 
This suggests, under the belief of unitarity, that perturbative methods are insufficient and so \textit{non-perturbative methods} \cite{Maldacena:2001kr} or new symmetry principles \cite{Hawking:2016,Addazi:2017} may have to be invoked to conclusively address the information loss paradox. Regarding the former approach, we recall that one of the most developed techniques is the \textit{Euclidean path-integral approach} \cite{Gibbons:1994cg,Hawking:2005kf}. 

In many situations, the non-perturbative approach through the Euclidean path-integral makes use of the concept of instantons.
We note, however, that Hawking radiation as it was originally derived is a perturbative phenomenon, while an instanton is a non-perturbative process. So at first glance the two notions seem incompatible. Nevertheless, we know that once gravity is taken into account, instantons can describe  phenomena associated with vacuum fluctuations that are consistent with that deduced from the perturbative quantum field theory \cite{Starobinsky:1986fx,Linde:1993xx}.

In this paper we will demonstrate that Hawking radiation is describable by Euclidean instantons. 
To accomplish this, we face two challenges.
\begin{itemize}
\item[1.] One issue is the complex-valued nature of instantons. As we will show below, this description requires \textit{complex-valued instantons} \cite{Halliwell:1989dy}. The Wick-rotation of time requires that all functions 
should be complexified. The meaning of such instantons has not been completely clarified yet.
Nevertheless, complex-valued instantons have been used in quantum cosmology 
and referred to as \textit{fuzzy instantons} \cite{Hartle:2007gi,Hwang:2011mp}, which have led to satisfactory results.
Thus we take the view that this approach is legitimate.
\item[2.] The other is the regularity of the horizon in Euclidean spacetime. If one chooses the Euclidean time period 
as the inverse of the Hawking temperature, then the matter distribution between the horizon and the infinity would cause a cusp at the horizon. Unless this cusp is treated by a proper method, one cannot assign a well-defined probability for such an instanton. Fortunately, however, there has been some progress in this issue, and we now have a good
reason to believe that the presence of a cusp may be appropriately regularized~\cite{Gregory:2013hja}.
Namely, such cusp instantons may be regarded as legitimate, which should contribute to the transition amplitude.
\end{itemize}
Adopting the above mentioned resolutions, we show that Hawking radiation 
can be described in terms of Euclidean instantons.
Our result implies that there may be additional situations where instantons can play a more useful role
than what we previously thought. If Hawking radiation can be described by the tunneling of instantons, it can cover not only the large mass limit, but also the small mass limit when the quantum effects are no more perturbative. This instanton approach may therefore provide a holistic methodology to shed lights on the eventual solution to the information loss paradox. We hope that this work may further promote this research direction~\cite{Sasaki:2014spa}.

This paper is organized as follows. In SEC.~\ref{sec:fro}, we construct scalar field instantons that include Hawking radiation. In SEC.~\ref{sec:haw}, we argue that these instantons cover the particle tunneling pictures, including Hartle-Hawking picture and Parikh-Wilczek picture.
%In SEC.~\ref{sec:bey}, we go beyond the perturbative regime and find meanings for the information loss problem.
Finally, in SEC.~\ref{sec:dis}, we summarize our discussions.

\section{\label{sec:fro}Construction of instantons}

\begin{figure}
\begin{center}
\includegraphics[scale=0.7]{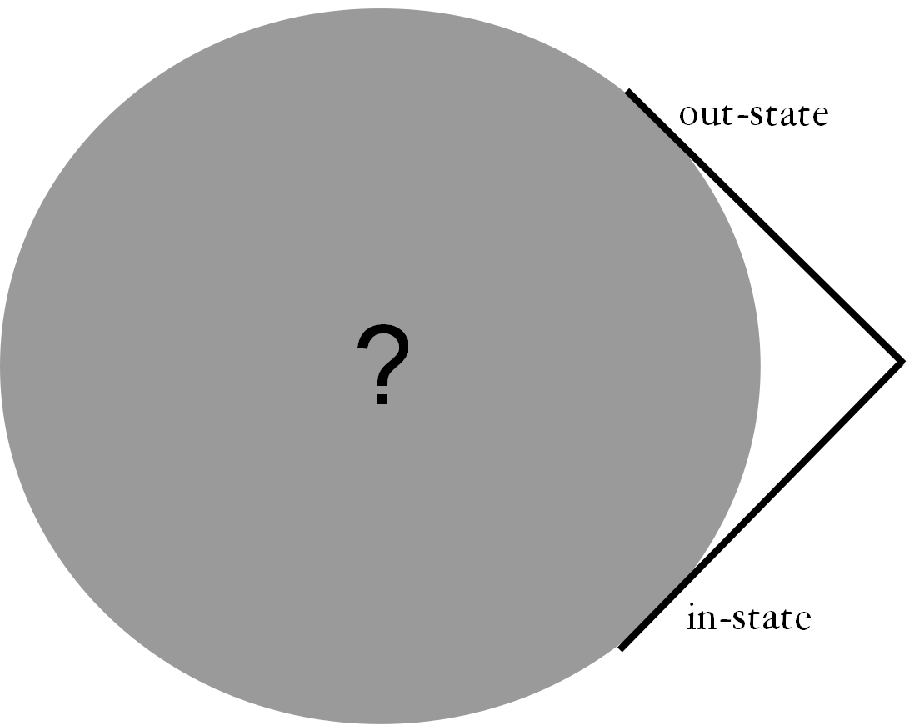}
\caption{\label{fig:conc1}We want to know the scattering amplitude from the in-state to the out-state.}
%\end{center}
%\end{figure}
%\begin{figure}
%\begin{center}
\includegraphics[scale=0.7]{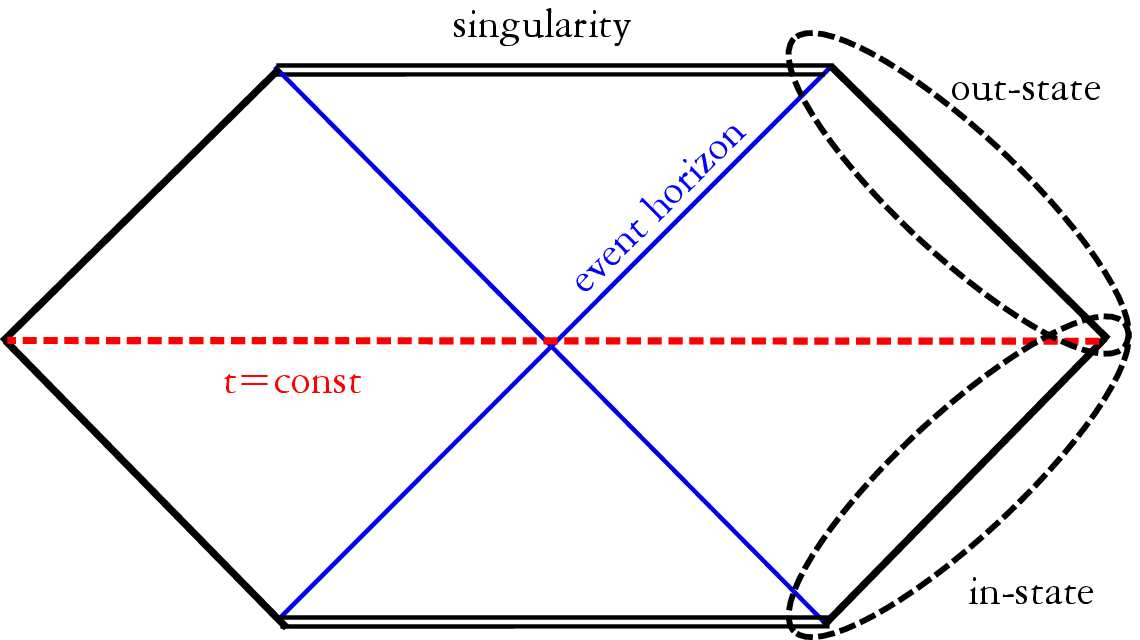}
\caption{\label{fig:conc2}The intermediate geometry will be well-approximated by the maximally extended geometry.}
\end{center}
\end{figure}

The scattering amplitude from an in-state (defined at the past null infinity, 
say $(h^{\mathrm{in}}_{ab}, \phi^{\mathrm{in}})$) to an out-state (defined at the future null infinity, say
 $(h^{\mathrm{out}}_{ab}, \phi^{\mathrm{out}})$) (Fig.~\ref{fig:conc1}) is formally defined by the path-integral,
\begin{eqnarray}
\Psi \left[ h^{\mathrm{out}}_{ab}, \phi^{\mathrm{out}}; h^{\mathrm{in}}_{ab}, \phi^{\mathrm{in}} \right] = \int \mathcal{D}g_{\mu\nu} \mathcal{D}\phi \;\; e^{i S[g_{\mu\nu},\phi]},
\end{eqnarray}
where we sum over all $g_{\mu\nu}$ and $\phi$ that connects $(h^{\mathrm{in}}_{ab}, \phi^{\mathrm{in}})$ and $(h^{\mathrm{out}}_{ab}, \phi^{\mathrm{out}})$. Although this integral is not easy to evaluate in practice,
extrapolating the flat space quantum field theory, we may assume that it can be evaluated by 
the analytic continuation to the Euclidean time $t = -i \tau$ \cite{Hartle:1983ai}:
\begin{eqnarray}
\Psi_{0} \left[ h^{\mathrm{out}}_{ab}, \phi^{\mathrm{out}}; h^{\mathrm{in}}_{ab}, \phi^{\mathrm{in}} \right] = \int \mathcal{D}g_{\mu\nu} \mathcal{D}\phi \;\; e^{- S_{\mathrm{E}}[g_{\mu\nu},\phi]},
\end{eqnarray}
which can be further approximated by invoking two assumptions. 
First, we assume a certain spacetime symmetry and restrict it to the mini-superspace. 
If the system includes a black hole, then the spherical symmetry is the simplest ansatz for the metric.
Second, we approximate the path-integral by using the steepest-descent approximation,
\begin{eqnarray}
\Psi_{0} \left[ h^{\mathrm{out}}_{ab}, \phi^{\mathrm{out}}; h^{\mathrm{in}}_{ab}, \phi^{\mathrm{in}} \right] \simeq \sum_{\mathrm{on-shell}} e^{- S_{\mathrm{E}}^{\mathrm{on-shell}}[g_{\mu\nu},\phi]},
\end{eqnarray}
where we sum over all on-shell solutions, i.e., the instantons, that connects 
$(h^{\mathrm{in}}_{ab}, \phi^{\mathrm{in}})$ and $(h^{\mathrm{out}}_{ab}, \phi^{\mathrm{out}})$\footnote{Note that we do not specify the exact form of the out-state, because for a given in-state $(h^{\mathrm{in}}_{ab}, \phi^{\mathrm{in}})$, we can construct various different contributions to the out-state $(h^{\mathrm{out}}_{ab}, \phi^{\mathrm{out}})$, as we will see in Subsection~\ref{sec:wick}. This flexibility is very important and this is the reason why we can finally recover the thermal distribution.}.

Now let us consider the meaning of on-shell solutions. First, we note that the Euclidean geometry 
connects the past null infinity, $(h^{\mathrm{in}}_{ab}, \phi^{\mathrm{in}})$, and the future null infinity, $(h^{\mathrm{out}}_{ab}, \phi^{\mathrm{out}})$.
For our purpose the initial condition may include the formation of the black hole, e.g., via a thin-shell collapse or
 a star collapse. While this adds complications to the problem, it is a classical process and hence as long as we study quantum 
 effects of a black hole, it may be reasonable to start with the maximally extended static black hole 
 (i.e., Schwarzschild) solution (Fig.~\ref{fig:conc2}); and consider instantons on the solution.

\begin{figure}
\begin{center}
\includegraphics[scale=0.6]{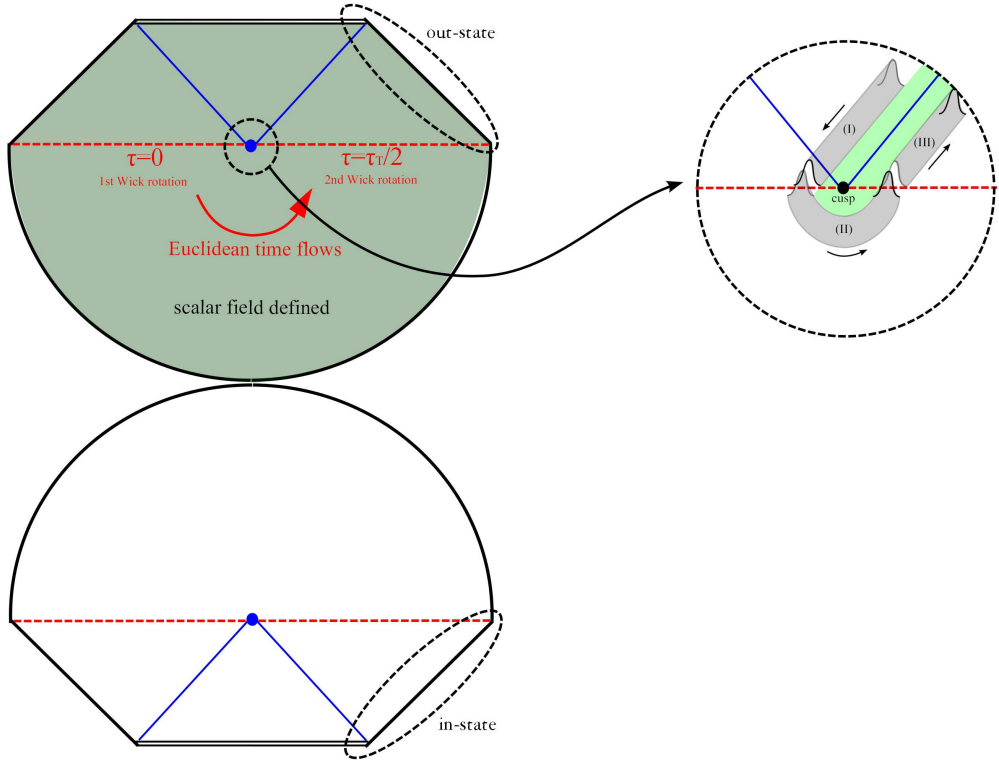}
\caption{\label{fig:conc3}A constant $t$ hypersurface will be analytically continued to the Euclidean manifold. The past (lower) part has no scalar field and the future (upper) part has a non-trivial scalar field solution. As we magnify around the symmetric point (Einstein-Rosen bridge), we can approximate that any solution is a superposition of in-going and out-going modes. We can choose a condition that there are only out-going modes. As we impose the reality condition at (III), the solution is complex-valued for (I) and (II). Since the out-going pulse has energy $\delta M$, the green colored region should have mass $M' = M - \delta M$. To cancel out the boundary term at infinity, we define the Euclidean time period as $\tau_{\mathrm{T}} = 8\pi M$. Therefore, the horizon becomes a cusp singularity.}
\end{center}
\end{figure}

The most trivial manifold for this setup is just the Lorentzian solution itself, but in this case the on-shell solutions have no quantum effects. The next simplest solution is the Euclidean Schwarzschild solution that is analytically continued at a certain time slice $\Sigma[t]$ (say, at $t=0$, where $t$ is the conventional Schwarzschild time coordinate). Then we can paste the Euclidean manifold to the past part as well as to the future part (Fig.~\ref{fig:conc3}). Although two Euclidean manifolds are apparently disconnected \cite{Chen:2015ibc}, such linkage between the initial state and the final state is considered consistent \cite{Gen:1999gi}.

Now we consider solutions with non-trivial field configurations on the manifold in Fig.~\ref{fig:conc3}. If we assume that the initial quantum state (in-state) is a vacuum, then at the on-shell level there should be no particles present in the past region. However, there exists a nonzero probability to find a non-vanishing field configuration in the future region (Upper of Fig.~\ref{fig:conc3}).

At this juncture we have a couple of questions:
\begin{itemize}
\item[1.] Since the solution is in general complex-valued, can one be sure that the solution satisfies 
the classicality, i.e., reality, condition at future null infinity?
\item[2.] Is the probability for each instanton solution consistent with Hawking's result?
\end{itemize}
Below we answer these questions,. We first define our model in Subsection~\ref{sec:model}, where we 
consider Einstein gravity with a massless scalar field. Then in Subsection~\ref{sec:wick}, we answer
 the first question and clarify that there exists a continuous family of non-trivial instanton solutions
 that are real at future null infinity. 
 Finally, in Subsection~\ref{sec:prob}, we answer the second question  by showing that such classicalized instantons indeed 
 correspond to the Hawking radiation.

\subsection{\label{sec:model}Model}

We consider Einstein gravity with a massless scalar field,
\begin{eqnarray}
S = \int dx^{4} \sqrt{-g} \left[ \frac{1}{16\pi} \mathcal{R} - \frac{1}{2} \left( \nabla \phi \right)^{2} \right] + \int_{\partial \mathcal{M}} \frac{\mathcal{K} - \mathcal{K}_{o}}{8\pi} \sqrt{-h} dx^{3},
\end{eqnarray}
where $\mathcal{R}$ is the Ricci scalar, $\mathcal{K}$ is the Gibbons-Hawking boundary term, and 
$\mathcal{K}_{o}$ is the corresponding boundary term for the periodically identified flat space~\cite{Gibbons:1976ue}.
Note that there is no potential term and so $\phi$ is a free scalar field. Since the Einstein equation gives 
$\mathcal{R} = 8\pi (\nabla \phi)^{2}$, the volume integration of the on-shell action vanishes.
Therefore for any field configuration, the on-shell Euclidean action becomes
\begin{eqnarray}\label{eq:prob}
S_{\mathrm{E}} = - \int_{\partial \mathcal{M}} \frac{\mathcal{K} - \mathcal{K}_{o}}{8\pi} \sqrt{+h} dx^{3} + \left( \mathrm{contribution\; at\; horizon} \right).
\end{eqnarray}
Note that the probability of a process mediated by this instanton is $P \sim e^{-2 B}$, where
\begin{eqnarray}
B = S_{\mathrm{E}}(\mathrm{solution}) - S_{\mathrm{E}}(\mathrm{background}).
\label{eq:bounce}
\end{eqnarray}

\subsection{\label{sec:wick}Two Wick rotations and the classicality of outgoing modes}

Let us construct an instanton solution as shown in Fig.~\ref{fig:conc3}. 
Namely, the initial state is a pure black hole, and the final state is a black hole
plus a scalar field.

Initially the spacetime is Schwarzschild,
\begin{eqnarray}
ds^{2} = - \left( 1 - \frac{2M}{R} \right) dt^{2} + \left( 1 - \frac{2M}{R} \right)^{-1} dR^{2} + R^{2} d\Omega^{2}\,,
\end{eqnarray}
where $-\infty<t\leq0$ on the physical side (i.e., the right-hand side of the spacetime in Fig.~\ref{fig:conc3}).
This is analytically continued to the Euclidean Schwarzschild solution at $t=0$,
\begin{align}
ds^2_{\mathrm{E}}= \left( 1 - \frac{2M}{R} \right) d\tau^{2} + \left( 1 - \frac{2M}{R} \right)^{-1} dR^{2} + R^{2} d\Omega^{2}\,,
\end{align}
where $\tau=it+const.$, and it is periodic with the period $\tau_{\mathrm{T}}=8 \pi M$. As shown in Fig.~\ref{fig:conc3},
half of this solution is matched to the initial Lorentzian Schwarzschild. For the sake of convenience,
we put $\tau=it-\tau_{\mathrm{T}}/2$ (equivalently, $t=-i(\tau+\tau_{\mathrm{T}}/2)$). 
So its range is $-\tau_{\mathrm{T}}/2 \leq \tau \leq 0$, and it matches to the unphysical side (i.e., the left-hand side 
in Fig.~\ref{fig:conc3}) of the same initial Lorentzian Schwarzschild solution at $\tau=0$.

Now we consider another Euclidean Schwarzschild solution but with a non-trivial scalar field configuration, with the range of the Euclidean time $0\leq\tau\leq\tau_{\mathrm{T}}/2$, which is eventually analytically continued to the Lorenztian solution for $t>0$ at $\tau=\tau_{\mathrm{T}}/2$ where $t=-i(\tau-\tau_{\mathrm{T}}/2)$. These two Euclidean solutions are disconnected, but the same formula Eq.~(\ref{eq:bounce}) can be consistently interpreted as having connected at infinity ($r\to\infty$), where the action $S_{\mathrm{E}}(\mathrm{background})$ corresponds to the first pure Schwarzschild instanton and $S_{\mathrm{E}}(\mathrm{solution})$ to the second one with a non-trivial field configuration.
 
To obtain this instanton, we first consider its analytic continued solution in the Lorentzian regime. The solution of the Klein-Gordon equation can be expressed as
\begin{eqnarray}
\phi = \sum_{\ell, m} a_{\ell m} \frac{f_{\ell}(t,r)}{R} Y_{\ell m}(\theta, \varphi),
\end{eqnarray}
where $R$ is the conventional Schwarzschild coordinate and $r$ is the tortoise coordinate,
\begin{eqnarray}
r = R + 2M \log \left| \frac{R}{2M} - 1 \right|,
\end{eqnarray}
and hence $-\infty < r < \infty$. The corresponding Klein-Gordon equation for each mode becomes
\begin{eqnarray}
\frac{\partial^{2} f_{\ell}}{\partial t^{2}} - \frac{\partial^{2} f_{\ell}}{\partial r^{2}} + \left( 1 - \frac{2M}{R} \right) \left( \frac{\ell(\ell+1)}{R^{2}} + \frac{2M}{R^{3}}\right) f_{\ell} = 0.
\end{eqnarray}
There is a potential barrier, but it disappears for $R \to 2M$ ($r \rightarrow - \infty$) as well as  
for $R \rightarrow \infty$ ($r \rightarrow \infty$).

Since we are interested in describing Hawking radiation, we focus on the field configurations 
in the near horizon limit ($R \to 2M$) in the Euclidean solution. Thus in the regions 
of the Lorentzian solution close to the matching surface, the waves are also confined
to the near horizon limit. In this limit the equation is well approximated by
\begin{eqnarray}
\frac{\partial^{2} f_{\ell}}{\partial t^{2}} - \frac{\partial^{2} f_{\ell}}{\partial r^{2}} = 0.
\end{eqnarray}
Since the equation is insensitive up to $\ell$, to avoid unnecessary complications, we further
focus on the $\ell=0$ modes, i.e., spherically symmetric configurations. In addition, since
we are interested in the radiation that is emitted to the future null infinity, we consider waves
propagating near the black hole (or future) horizon. Therefore, 
\begin{align}
\phi=\frac{f(t,r)}{R(r)}
=\frac{1}{\sqrt{2\pi}} \int_{-\infty}^{\infty} \frac{d\omega}{R(r)} a_{\omega}^{\rm out} e^{-i \omega (t-r)}\,.
\end{align}

Now we match this to the Euclidean solution through the time slice at $t=0$. 
With $t=-i(\tau-\tau_{\mathrm{T}}/2)$, the above gives
\begin{align}
\phi=\frac{1}{\sqrt{2\pi}} \int_{-\infty}^{\infty} \frac{d\omega}{R(r)} a_{\omega}^{\rm out} e^{\omega\tau_T/2}
e^{-\omega \tau +i\omega r }\,.
\label{eq:eucsol}
\end{align}
This is matched to the Lorentzian solution at $\tau=0$ to the unphysical, $t<0$, region of the solution.
\begin{align}
\phi=\frac{1}{\sqrt{2\pi}} \int_{-\infty}^{\infty} \frac{d\omega}{R(r)} a_{\omega}^{\rm in}e^{-i \omega (t-r)}\,;
\quad a_\omega^{\rm in}=a_\omega^{\rm out}e^{\omega\tau_T/2}.
\end{align}
Now since $t$ runs backward in time on the unphysical side of the solution, this describes waves
propagating out of the black hole in the past direction. Thus seeing from the physical side, they
are the negative energy waves that compensate the energy carried out by waves propagating along
and outside the black hole horizon.

Keeping the above discussion in mind, now we impose the physical requirements. 
First, we demand the asymptotic classicality, namely the out-going mode should contain only real particles. 
Second, we impose the energy conservation law. Thus if there is an out-going energy flux, then the black hole mass
(Misner-Sharp mass) should be smaller than the ADM mass at infinity.

The reality condition implies
\begin{eqnarray}
a_{\omega}^{{\rm out}*} = a_{-\omega}^{\rm out}.
\end{eqnarray}
Note that this implies that the solution in the Euclidean time, Eq.~(\ref{eq:eucsol}), is complex:
\begin{align}
\phi^*\neq\phi\quad\mbox{for real $\tau$}\,.
\end{align}
As discussed in the Introduction, we take the view that complex-valued instantons
are legitimate solutions that contribute to the transition amplitude.

To track the energy carried by the radiation, we express the field as
\begin{equation}
\phi(t, r) = \frac{1}{\sqrt{2\pi}R(r)} \int_{0}^{\infty} \frac{d\omega}{\sqrt{2\omega}} 
\left[A_{\omega} e^{- i \omega (t - r)} + A^{*}_{\omega} e^{i \omega (t - r)} \right],
\end{equation}
where $A_{\omega} = \sqrt{2\omega} a_{\omega}^{\rm out}$. 
Since the Hamiltonian, $H$, around the horizon is
\begin{eqnarray}
H = \delta M \propto \int_{0}^{\infty} d\omega \omega |A_{\omega}|^{2}\,,
\end{eqnarray}
the number of out-going particles is $N_{\omega} \propto |A_{\omega}|^{2}$. 
Hence, the Misner-Sharp mass at the horizon should be $M' = M - \delta M$ due to the energy conservation.

\subsection{\label{sec:prob}Probabilities}

Now we proceed to compute the probability of the process mediated by the instanton.
Based on Eq.~(\ref{eq:prob}), since there is no contribution from the volume integration, the probability interpretation is
the same as that of the thermal thin-shell instanton \cite{Garriga:2004nm}. The only contributions are from 
the boundary terms; one from the boundary at infinity and the other at the horizon.

Regarding the boundary term at infinity, since we consider matching the two disconnected instantons 
at infinity,  naturally they cancel each other. Thus the only possible contribution is from the horizon.
If there were no cusp, there would be no contribution from the horizon either. But in our case
we do have a cusp because of the difference between the period associated with
the black hole in the out state and the period fixed at infinity. Thus we expect to have
some contribution from the cusp singularity after some proper regularization~\cite{Chen:2017suz}.
To regularize the cusp, we can apply the same procedure used for thin-shell 
instantons~\cite{Gregory:2013hja}. We obtain
\begin{eqnarray}
2B = \frac{\mathcal{A}}{4} - \frac{\mathcal{A}'}{4} = 4\pi \left( M^{2} -M'^{2} \right),
\end{eqnarray}
where $\mathcal{A}$ and $M$ are the areal radius and mass of the initial black hole, respectively,
while the primed ($'$) quantities denote those for the final black hole. One can calculate the probability for 
the same process by using the Hamiltonian approach \cite{Fischler:1989se} and can obtain the same 
result \cite{Chen:2015ibc}. Hence, we obtain a transition probability that depends only on the area change.

To evaluate the probability, we first consider an emission of a single quantum with energy $\omega$ of the order of $\mathcal{O}(M^{-1})$. 
For a large black hole with $M\gg 1$, we have $M' = M - \omega$ with $\omega \ll M$.
In this limit, we have
\begin{eqnarray}
2B = 8\pi M \omega\,.
\end{eqnarray}
Thus the exponent is perfectly consistent with Hawking radiation, with
the Hawking temperature identified as $T_{\mathrm{H}}=(8\pi M)^{-1}$. 
Following the usual procedure \cite{Hackworth:2004xb}, the total partition function 
may be evaluated as
\begin{eqnarray}
Z = Z_{0} \left( 1 + \frac{i}{2} \Omega \mathcal{T} K e^{-2B} \right),
\end{eqnarray}
where $\Omega$ is the volume, $\mathcal{T}$ is the time, $K$ is determined from the 
perturbations around the instanton, and 
$Z_{0} = e^{-2 S_{\mathrm{E}}(\mathrm{background})} [\mathrm{det} S_{\mathrm{E}}''(\mathrm{background})]^{-1/2}$. 
As usual this gives the decay rate, 
\begin{eqnarray}
\Gamma = 2 \lim_{\Omega,\mathcal{T} \rightarrow \infty} 
\left[\frac{\mathrm{Im} \log Z}{\Omega \mathcal{T}}\right] \simeq K e^{-2B}.
\end{eqnarray}
Summing over all possible numbers of quanta ($\omega$, $2\omega$, $3\omega$, ...), 
\begin{align}
Z = Z_{0} \left( 1 + \frac{i}{2} \Omega \mathcal{T} K \sum_{n=1}^{\infty} e^{-2nB} \right),
\end{align}
we then readily recover the Planck distribution with $T = 1/8\pi M$ as we advertised \cite{Banerjee:2009wb}.

\subsection{Analogy to the Bogoliubov transformation}

Our approach is not the same as the canonical approach based on the Bogoliubov transformation \cite{Hawking:1974sw,Birrell:1982ix}. However, we can appreciate their analogy through the following argument.

The scalar field corresponds to particles for $\omega > 0$ and antiparticles for $\omega < 0$. Then before the Wick-rotation, the solution looks like
\begin{equation}
\phi(t, r) = \frac{1}{\sqrt{2\pi}R(r)} \int_{0}^{\infty} d\omega \left( a_{\omega} e^{- i \omega (t - r)} + a_{-\omega} e^{i \omega (t - r)} \right).
\end{equation}
Note that the coordinate time of the left side of the Penrose diagram flows oppositely. In order to identify this with the causal time, one can substitute $t \rightarrow - t$, and hence,
\begin{equation}
\phi(t, r) = \frac{1}{\sqrt{2\pi}R(r)} \int_{0}^{\infty} d\omega \left( a_{-\omega} e^{- i \omega (t + r)} + a_{\omega} e^{i \omega (t + r)} \right).
\end{equation}
Now we identify this with the in-state of the Bogoliubov transformation, where $a_{-\omega}$ corresponds the particle modes and $a_{\omega}$ corresponds the antiparticle modes. In order to avoid the confusion, we redefine the mode functions such that
\begin{equation}
\phi(t, r) = \frac{1}{\sqrt{2\pi}R(r)} \int_{0}^{\infty} d\omega \left( \tilde{b}_{\omega} e^{- i \omega (t + r)} + \tilde{b}_{-\omega} e^{i \omega (t + r)} \right),
\end{equation}
by identifying $a_{-\omega} = \tilde{b}_{\omega}$ and $a_{\omega} = \tilde{b}_{-\omega}$.

\begin{figure}
\begin{center}
\includegraphics[scale=0.45]{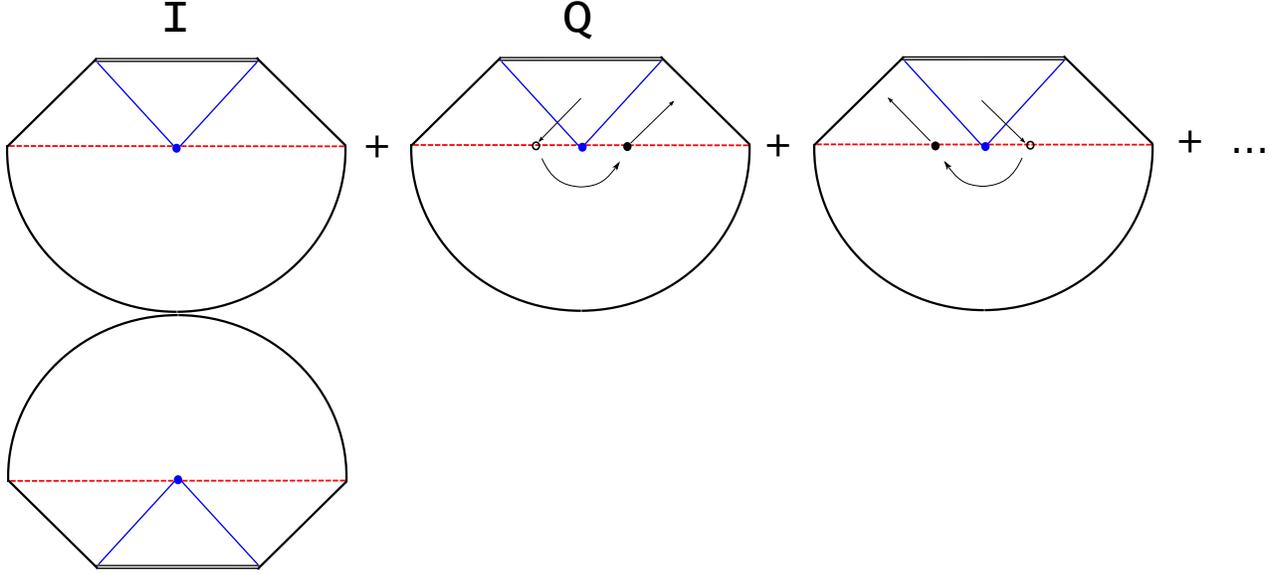}
\caption{\label{fig:conc33}As we sum over all instantons, $\mathcal{I}$ (Schwarzschild without scalar fields) corresponds the trivial process. There can be various instantons, but we only select instantons that has out-going modes, e.g., $\mathcal{Q}$. If we sum $\mathcal{I}$ and $\mathcal{Q}$, one can recover the Bogoliubov matrix at least in the heuristic level.}
\end{center}
\end{figure}

After the Wick-rotation, the field becomes
\begin{equation}
\phi(t, r) = \frac{1}{\sqrt{2\pi}R(r)} \int_{0}^{\infty} d\omega \left( b_{\omega} e^{- i \omega (t - r)} + b^{*}_{\omega} e^{i \omega (t - r)} \right).
\end{equation}
From the investigations of the previous subsection, we obtain a linear transformation between ($\tilde{b}_{\omega}$, $\tilde{b}_{-\omega}$) and ($b_{\omega}$, $b^{*}_{\omega}$) such that
\begin{eqnarray}
\left( \begin{array}{c}
b_{\omega} \\
b^{*}_{\omega}  
\end{array} \right) =
\left( \begin{array}{cc}
0 & e^{-\omega \tau_{\mathrm{T}}/2} \\
e^{+\omega \tau_{\mathrm{T}}/2} & 0
\end{array} \right)
\left( \begin{array}{c}
\tilde{b}_{\omega} \\
\tilde{b}_{-\omega}  
\end{array} \right).
\end{eqnarray}
The reality condition after the Wick-rotation requires that $b_{\omega} = b^{*}_{\omega}$, i.e., $a_{-\omega} = e^{-\omega \tau_{\mathrm{T}}} a^{*}_{\omega}$. By using this relation, we can present the equivalent transformation such that
\begin{eqnarray}
\left( \begin{array}{c}
b_{\omega} \\
b^{*}_{\omega}  
\end{array} \right) = \mathcal{Q} \left( \begin{array}{c}
\tilde{b}_{\omega} \\
\tilde{b}^{*}_{\omega}  
\end{array} \right) \equiv
\left( \begin{array}{cc}
0 & e^{-\omega \tau_{\mathrm{T}}/2} \\
e^{-\omega \tau_{\mathrm{T}}/2} & 0
\end{array} \right)
\left( \begin{array}{c}
\tilde{b}_{\omega} \\
\tilde{b}^{*}_{\omega}  
\end{array} \right).
\end{eqnarray}

Now let us find an analogy with the usual Bogoliubov transformation. If there is no mixing between different frequencies, then the Bogoliubov transformation from the in-going modes to the out-going modes will be
\begin{eqnarray}
\left( \begin{array}{c}
b_{\omega} \\
b^{*}_{\omega}  
\end{array} \right) = \mathcal{U} \left( \begin{array}{c}
\tilde{b}_{\omega} \\
\tilde{b}^{*}_{\omega}  
\end{array} \right) =
\left( \begin{array}{cc}
\alpha_{\omega} & \beta^{*}_{\omega} \\
\beta_{\omega} & \alpha^{*}_{\omega} 
\end{array} \right)
\left( \begin{array}{c}
\tilde{b}_{\omega} \\
\tilde{b}^{*}_{\omega}  
\end{array} \right)
\end{eqnarray}
with the normalization condition $|\alpha_{\omega}|^{2} - |\beta_{\omega}|^{2} = 1$. This transformation matrix should include all contributions from different paths and will be approximated by (Fig.~\ref{fig:conc33})
\begin{eqnarray}
\mathcal{U} \simeq \mathcal{I} + \mathcal{Q},
\end{eqnarray}
where $\mathcal{I}$ is the identity matrix (which represents the trivial classical mode propagation) and $\mathcal{Q}$ is the contribution from the Euclidean path-integral that is approximated by instantons. Then, the Bogoliubov matrix becomes
\begin{eqnarray}
\mathcal{U} = \frac{1}{\sqrt{N}}
\left( \begin{array}{cc}
1 & e^{-\omega \tau_{\mathrm{T}}/2} \\
e^{-\omega \tau_{\mathrm{T}}/2} & 1 
\end{array} \right),
\end{eqnarray}
where $N = 1 - e^{-\omega \tau_{\mathrm{T}}}$ is the normalization factor. Finally, we obtain the Hawking-like relation such that
\begin{eqnarray}
|\beta_{\omega}|^{2} = |t_{\omega}|^{2} \frac{e^{-\omega \tau_{\mathrm{T}}}}{1-e^{-\omega \tau_{\mathrm{T}}}},
\end{eqnarray}
where $t_{\omega}$ is the transmission coefficient for the modes that propagate in large $r$ region.

This is of course not a bona fide quantum field theoretical derivation, since we did not regard all coefficients as real operators. Rather, we heuristically identify the relation between mode coefficients. However, it is worthwhile to see that our results interestingly reproduce that of Hawking radiation.

\section{\label{sec:haw}Hawking radiation revisited}

Our result reveals that instanton approach can indeed recapitulate the Hawking radiation with a thermal spectrum. Traditionally, there have been at least three approaches to Hawking radiation.

First, one can derive Hawking radiation by using the Bogoliubov transformation, 
which we have checked and demonstrated the consistency between the two approaches in the previous subsection,
at least heuristically.

Second, one can derive the evaporation of a black hole by using the renormalized energy-momentum
 tensor~\cite{Birrell:1982ix}. In order to obtain the renormalized energy-momentum tensor, one needs to invoke 
 a regularization method. The simplest example of exact solution was obtained in $2$-dimensions \cite{Davies:1976ei}, 
 which may be extended to higher dimensional cases using
 the $S$-wave approximation \cite{Hwang:2010im}. It has been shown that the Euclidean approach may reproduce the approximate forms of the regularized 
 energy-momentum tensors \cite{Page:1982fm}.

Third, one can derive it by using the concept of tunneling \cite{Hartle:1976tp,Parikh:1999mf}. 
One can interpret Hawking emission as the tunneling of a particle with energy $\omega$, where the decay rate
$\Gamma \sim e^{- \omega/T}$ and the temperature $T$ can be estimated. In this tunneling approach, one obtains the 
Boltzmann distribution. After summing over all quanta, one recovers the Planck 
distribution \cite{Banerjee:2009wb}.

In this section we further discuss this tunneling picture \cite{Hartle:1976tp,Parikh:1999mf},
 since the tunneling approach has a direct correspondence to the instanton picture.

\begin{figure}
\begin{center}
\includegraphics[scale=0.5]{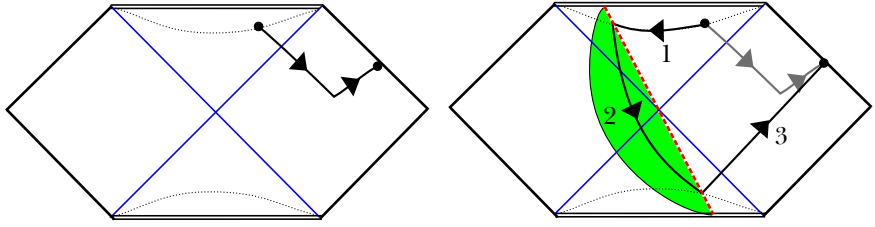}
\caption{\label{fig:cycles4}Summarization of the Hartle-Hawking picture. Left: Particle tunneling from inside to outside. Right: An equivalent path according to Hartle and Hawking. Path (1) is along the constant $r$ surface, that does not contribute to the propagator, due to the time-translational invariance. Path (2) relies on the Euclidean analytic continuation. Due to the time-reversal symmetry, Path (3) can be interpreted to contribute to the absorption process.}
\end{center}
\end{figure}

\subsection{Hartle-Hawking method}

Although the original derivation by Hawking used the Bogoliubov transformation
method~\cite{Hawking:1974sw}, soon an explanation based on the pair-creation of
particles was given by Hartle and Hawking~\cite{Hartle:1976tp}. 
They considered the path-integral of a particle that starts 
 from inside the black hole, goes \textit{backward} in time, reaches outside the event horizon, and then goes 
 forward in time to the asymptotic future infinity (Left of Fig.~\ref{fig:cycles4}). Although such a trajectory is not allowed 
 in classical physics, it is possible in quantum mechanics.

Hartle and Hawking calculated the decay rate by taking the following logical steps (Right of Fig.~\ref{fig:cycles4}).
\begin{description}
\item[--]Step 1: For a black hole formed by gravitational collapse, the tunneling process
can be described by the analytic continuation of the static black hole solution.
\item[--]Step 2: The process to be computed is the tunneling in the left diagram in Fig.~\ref{fig:cycles4}. 
The emission probability can be calculated from the propagator, i.e., the path-integral, of the particle. 
One then invokes the Euclidean analytic continuation to find out an equivalent integration.
\item[--]Step 3: In order to calculate the path-integral or the propagator, one can rely on the following three properties.
\begin{enumerate}
\item The path-integral is \textit{analytic} up to the analytic continuation over the Euclidean time. Hence, the dominant contribution only depends on the initial point and the final point (large dots in the Left diagram in Fig.~\ref{fig:cycles4}).
\item The path-integral has the \textit{time-translation symmetry} for a given $r$. Therefore, Path (1) in Fig.~\ref{fig:cycles4} does not contribute to the propagator.
\item The path-integral has the \textit{time-reversal symmetry}. Hence, Path (3) in Fig.~\ref{fig:cycles4} is equivalent to the absorption probability for a given particle.
\end{enumerate}
\item[--]Step 4: Hartle and Hawking chose the equivalent path as that indicated in the Right diagram in Fig.~\ref{fig:cycles4}. 
Therefore, the probability is contributed by two parts: the Boltzmann factor from the Euclidean 
time evolution (a half period of the Euclidean manifold) and the absorption probability of the particle.
\end{description}

\begin{figure}
\begin{center}
\includegraphics[scale=0.5]{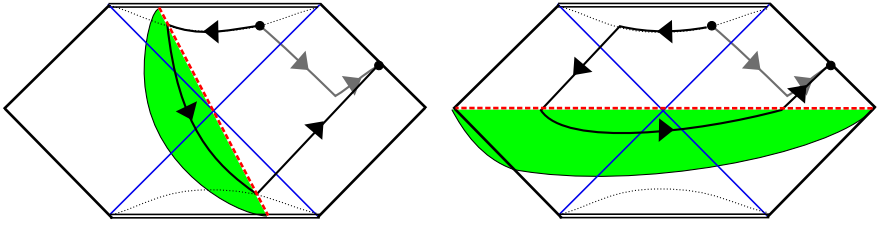}
\caption{\label{fig:cycles5}Left: The calculation trick by Hartle and Hawking. Right: Due to analyticity, we can choose different path. Then this will give the same results of the instanton approach.}
\end{center}
\end{figure}

However, this approach has several problems. First, the Euclidean analytic continuation cannot form
 a Euclidean manifold with the signatures $(+,+,+,+)$; rather, it forms $(+,+,-,-)$. Since there is no 
 well-defined manifold for such a signature, it is not easy to impose back-reactions; Hartle and 
 Hawking's description therefore only holds either for emissions with negligible energy $\omega \ll M$, or for a perfect thermal equilibrium system
 with negligible black hole mass decrease \cite{Israel:1976ur}. One therefore needs find an equivalent steepest-descent method that is applicable to the Euclidean manifold that include processes with non-negligible emission energy $\omega$. 
 Note that it is conceptually easy to see that the Hartle-Hawking approach is analytically equivalent to the
 instanton approach (Fig.~\ref{fig:cycles5}), while the latter approach only invokes the Euclidean manifold 
 that satisfies the energy conservation consistently.

\begin{figure}
\begin{center}
\includegraphics[scale=0.5]{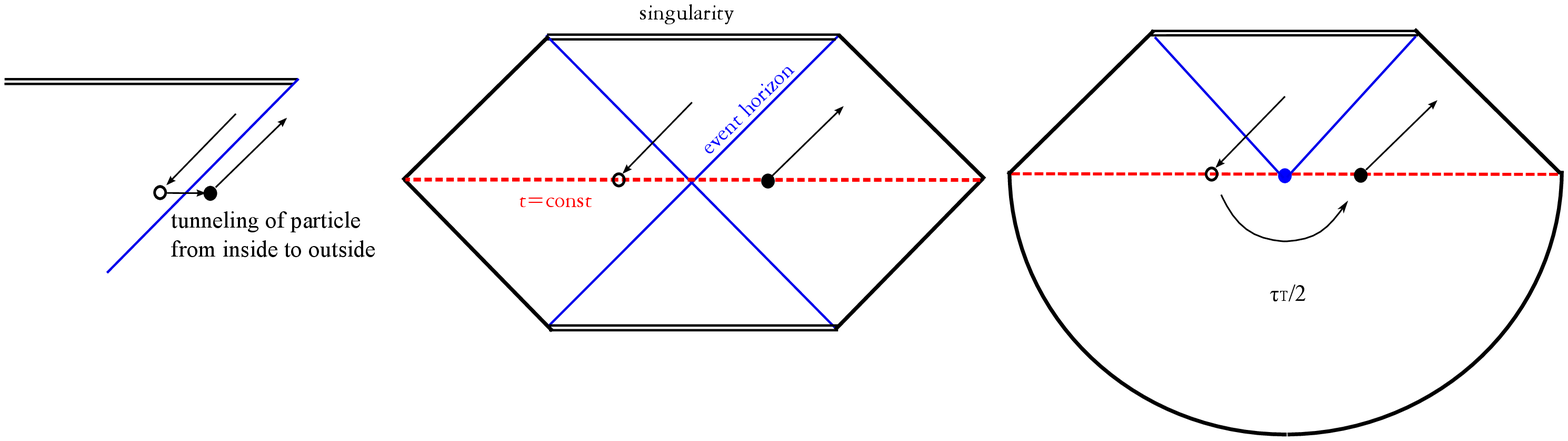}
\caption{\label{fig:for_note2}Left: The Parikh-Wilczek method describes tunneling from inside to outside the event horizon. Middle: The important contribution is tunneling from left (white dot) to right side (black dot) of the whole causal structure. Right: Two particles can be connected by the Euclidean manifold outside the horizon.}
\end{center}
\end{figure}

\subsection{Parikh-Wilczek method}

Parikh and Wilczek \cite{Parikh:1999mf} developed a more direct description of this process without 
using the Euclidean analytic continuation. In this approach the Hawking evaporation is treated as the tunnelling of a particle from inside to outside the horizon 
(Left of Fig.~\ref{fig:for_note2}). When we consider tunneling of particles with energy $\omega$, 
the decay rate is
\begin{eqnarray}
\Gamma \sim e^{- 2 \mathrm{Im} S} \sim e^{- \omega/T}
\end{eqnarray}
where $T$ is the temperature and $S$ is the Lorentzian action of the particle.
Such a tunneling between two null geodesics (one inside and the other outside the event horizon) is 
 not allowed classically. Quantum mechanically, however, the action of the particle can acquire an imaginary part such that
\begin{eqnarray}
\mathrm{Im} S = \mathrm{Im} \int_{r_{\mathrm{in}}}^{r_{\mathrm{out}}} p_{r} dr 
\simeq \omega \frac{\tau_{\mathrm{T}}}{2},
\end{eqnarray}
where $\tau_{\mathrm{T}} = 1/T$ is the period of the Euclidean time.

It is interesting to compare this with the instanton picture. The dominant contribution of the tunneling
 between two null geodesics is equivalent to the tunneling between the left side and the right side
  of the Einstein-Rosen bridge (Middle of Fig.~\ref{fig:for_note2}). 
  These two geodesics can be connected by the Euclidean manifold (Right of Fig.~\ref{fig:for_note2}).

We emphasize that our analysis in terms of instantons is consistent with but very different from the Parikh-Wilczek method, which deals with the tunneling of a particle through the black hole horizon and the tunneling rate is given by
computing the action of the particle that acquires an imaginary part when crossing the horizon. In contrast, we deal with a scalar field and construct \textit{complex} instantons that describe the tunneling of a black hole with no scalar
field to a black hole with an outgoing scalar field at future null infinity as Hawking radiation. The physical pictures between the two interpretations are therefore also very different.

\section{\label{sec:dis}Discussion}

In this paper, we derived Hawking radiation using the instanton approach. We interpret the Hawking radiation process
as the sum over all possible paths of instantons with certain physical constraints such as the energy conservation and the asymptotic classicality. The probability so obtained is consistent with Hawking's result. It is thus reasonable to think that 
the \textit{black hole evaporation can be regarded as the sum of all possible trajectories of a series of instantons}. Even though we discussed
 $4$-dimensional massless scalar field cases, it should be very easy to extend it to other dimensions and other fields.

Our result is going one step further from Hartle-Hawking and Parikh-Wilczek. First of all, our approach used a scalar field rather than particle trajectories, where the previous tunneling descriptions have been limited to the particle-level approaches. In addition, it is important to reemphasize that the instanton method is applicable to much broader areas
than Hawking radiation. For example, if we increase the amplitude of the field value, then the instanton is no more Hawking radiation but becomes a kind of non-perturbative effects. In other words, not only can it cover perturbative effects (Hawking radiation) but also non-perturbative 
effects that may even explain, for example, the transition to a trivial geometry \cite{Hawking:2005kf}. Although in that case their transition probabilities are suppressed, they have nevertheless non-zero probabilities.

These non-perturbative effects may in principle be invoked to resolve the information loss problem, especially in the $\omega \simeq M$ limit. The present paper reasonably indicates that the existence of non-perturbative channels are indeed very generic for various gravity models, since they manifest themselves even in the case of a free scalar field. It is not so surprising that these instantons can provide us insights into the information loss paradox. Since instantons are related to the quantum gravitational wave function, such a formulation can cover not only perturbative phenomena but also non-perturbative phenomena \cite{Maldacena:2001kr}. If one takes into account all non-perturbative effects that allow for the transition toward a trivial geometry, then the entire wave function is no more semi-classical and one should be able to see new features of the information loss problem beyond what has been considered. In addition, non-perturbative contributions should dominate the late stage of the black hole evolution; then, instantons may provide important clues toward the understanding of the late stage black hole evaporation. This, however, is beyond the scope of the present paper and we postpone the more detailed discussion of it to a future publication.

In this work there are several directions that need to be improved. We considered instantons only near the horizon; also, 
we assumed that the metric back-reactions are negligible. These assumptions are reasonable for $\omega \ll M$, 
but if $\omega \lesssim M$, it becomes more complicated. The probability interpretation should be the same since 
there is no contribution from the volume integration of the action; however, the reality condition for both the field 
and the metric at infinity is not clear. This should be confirmed by more detailed numerical investigations. In addition, it may be possible to apply our instanton method to investigate the radiation for an accelerating observer \cite{Unruh:1976db}, the acoustic radiation in a supersonic fluid \cite{Unruh:1980cg}, or the radiation from a moving mirror \cite{Davies:1976hi}. This is an interesting and challenging topic for future investigations. Furthermore, if one extends our description beyond the steepest-descent approximation, one may be able to obtain more exact descriptions about quantum gravitational behaviors around black holes.

Although the Euclidean approach has several limitations, we believe that this is the right way of thinking and we hope that this instanton approach would shed lights on the information loss problem.

\section*{Acknowledgment}

DY would like to thank Takahiro Tanaka, Piljin Yi, and Guillem Dom\`enech for critical comments when some ideas of this paper were primitive. DY was supported by the Korea Ministry of Education, Science and Technology, Gyeongsangbuk-Do and Pohang City for Independent Junior Research Groups at the Asia Pacific Center for Theoretical Physics and the National Research Foundation of Korea (Grant No.: 2018R1D1A1B07049126).
This work was supported by the JSPS KAKENHI Nos. 15H05888 and 15K21733,
by World Premier International Research Center Initiative (WPI Initiative), MEXT, Japan,
and by MOST 107-2811-M-002-042, Taiwan.

\end{document}